# Twisted Radiation by Electrons in Spiral Motion

M. Katoh[1,2*], M. Fujimoto[1,2], N. S. Mirian[1], T. Konomi[1,2], Y. Taira[3], T. Kaneyasu[4], M. Hosaka[5], N. Yamamoto[5], A. Mochihashi[5], Y. Takashima[5], K. Kuroda[6], A. Miyamoto[7], K. Miyamoto[7], S. Sasaki[7]

[1]Institute for Molecular Science, National Institutes of Natural Sciences, Okazaki, 444-8585 Japan
[2]Sokendai (The Graduate University for Advanced Studies), Okazaki, 444-8585 Japan
[3]National Institute of Advanced Industrial Science and Technology (AIST), Tsukuba, 305-8565 Japan
[4]Saga Light Source, Tosu, 841-0005 Japan
[5]Nagoya University, Nagoya, 464-0814 Japan
[6]University of Tokyo, Kashiwa, 277-0882 Japan
[7]Hiroshima University, Higashi-Hiroshima, 739-0046 Japan

**ABSTRACT**

We theoretically show that a single free electron in circular/spiral motion radiates an electromagnetic wave possessing helical phase structure and carrying orbital angular momentum. We experimentally demonstrate it by double-slit diffraction on radiation from relativistic electrons in spiral motion. We show that twisted photons should be created naturally by cyclotron/synchrotron radiations or Compton scatterings in various situations in cosmic space. We propose promising laboratory vortex photon sources in various wavelengths ranging from radio wave to gamma-rays.

Twisted photons possessing helical wave front carry orbital angular momentum other than well-known spin angular momentum[1]. Such photons have been investigated theoretically and experimentally, in particular, towards applications in the information technologies, the nanotechnologies and the imaging technologies[2,3]. Their interactions with nuclei, atoms, molecules, materials or plasma have been also being explored theoretically rather than experimentally[2,3,4], that might be due to the lack of the laboratory photon sources other than in the laser wave length range where twisted photon beams can be readily obtained by using conventional optical devices. Twisted photons in nature, particularly in astrophysics, have been reviewed[5,6]. However, the authors did not discuss the sources of twisted photons explicitly because they could not find promising radiation processes. Instead, they discussed detection schemes of twisted photons in astronomical observations. Other authors proposed twisted radiation from rotating black halls or inhomogeneous interstellar media[7,8], however, they discussed modification of radiation rather than radiation processes itself.

In this paper, we show for the first time that a single free electron in circular/spiral motion emits twisted photons. This is one of the most fundamental radiation processes by a free electron and is the basis of synchrotron/cyclotron radiations or Compton scattering, which play important roles in various situations in nature or in laboratories, such as around magnetized neutron stars, in supernova explosions with magnetic fields, in nuclear fusion plasma, electron accelerators and so on. The radiations from such electrons cover the entire range of wave lengths, from radio wave to gamma-rays, depending on the physical parameters such as the electron energy or radius of motion. They should play unexplored, important roles with their angular momentum, in nature and in laboratories. Moreover, this radiation process can be the basis of laboratory vortex photon sources in the entire wavelength range, which will open new research opportunities.

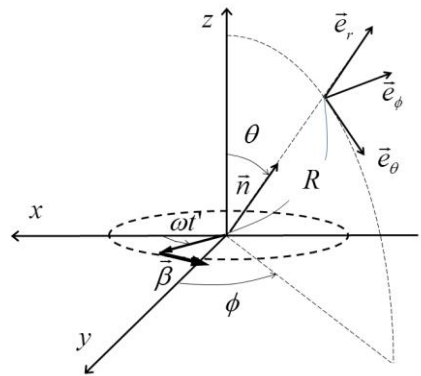

Figure 1. Coordinate system. The electron is rotating in the x-y plane around the origin with an initial position on the x-axis. The azimuthal angle of the position of the observer is measured from the y-axis. The observer frame is defined as a spherical coordinate.

Radiation generated by electrons in circular motion was first studied by O. Heaviside in

1904[9] and has since been addressed in many scientific reports[10, 11] and textbooks[12, 13]. However, we have been unable to find any discussion of its phase or wave front structure. In the following, we derive an analytic expression that contains a phase term and explain intuitively the mechanism which produces the twisting of the field. We assume that the electron trajectory draws a circle in a plane perpendicular to the z-axis. The radiation field is given by the following formulae, which are directly derived from the Liénard–Wiechert potentials[12]:

$$\vec{E}(t) = \frac{e}{cR} \frac{\vec{n} \times \left\{ (\vec{n} - \vec{\beta}) \times \dot{\vec{\beta}} \right\}}{(1 - \vec{n} \cdot \vec{\beta})^3} \bigg|_{t'} \qquad (1)$$

$$\vec{H}(t) = \vec{n} \times \vec{E} \big|_{t'} \qquad (2)$$

$$t' = t - \frac{R}{c} \qquad (3)$$

Here, $\vec{E}(t)$ and $\vec{H}(t)$ are the electric and magnetic fields, respectively, $c$ is the speed of light, $e$ is the elemental charge, $\vec{n}$ is the unit vector, $R$ is the distance from the origin to the observer (see Fig. 1), $\vec{\beta}$ is the electron velocity normalised by the velocity of light, and $\omega$ is the electron angular velocity. The right sides of Eqs. (1) and (2) represent the emitter time $t'$, which is related to the observer time $t$ by Eq. (3). An expression for the electric field in the observer frame is obtained by inserting adequate expressions for the vectors in Eq. (1):

$$\vec{E}(R,\theta,\phi,t) = \frac{e}{cR} \frac{\beta\omega}{\{1 - \beta \sin\theta \cos(\omega t' - \phi)\}^3} \left[ \cos\theta \sin(\omega t' - \phi) \vec{e}_\theta - \{\cos(\omega t' - \phi) - \beta \sin\theta\} \vec{e}_\phi \right] \qquad (4)$$

Because the magnetic field can be obtained using Eq. (2), we do not show its explicit form.

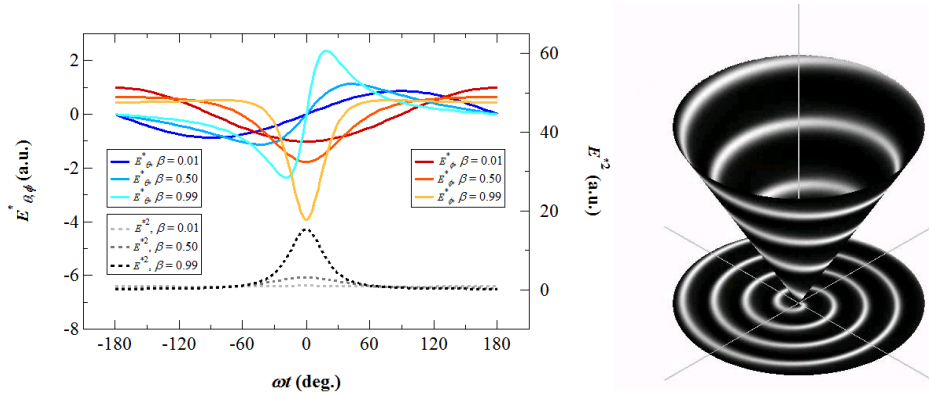

Figure 2. Left: Waveform of the electric field propagating towards the polar angle ($\theta$) of 30° and the azimuthal angle ($\phi$) of 0° for a range of electron velocities, $\beta$. Blue lines

represent the $\theta$ components, and red lines represent the $\phi$ components (see Fig. 1). Black dotted lines show the field intensities, given as the square summation of the electric field components. To emphasise the change in the waveform, the electric fields are divided by $\beta$ in the calculation (see Eq. (4)). Right: Radiation field intensity propagating towards 30° from the z-axis and its projection on the x-y plane. The electron velocity $\beta$ is 0.5. The brightness is the magnitude.

The waveforms of the radiation fields calculated using Eq. (4) are shown in Fig. 2. As the electron velocity increases, the sinusoidal waveforms become distorted and develop kinks around the phase, corresponding to the time when the instantaneous electron motion is directed to the observer (Fig. 1). The deformed wave can be decomposed into harmonic components with frequencies that are integer multiples of $\omega$ [9].

The spatial distribution of the propagating electric field is also shown in Fig. 2. The relativistic kink where the field is strengthened is distributed as a spiral, suggesting that the phase of the higher harmonic components also depends on the azimuthal angle. We decompose Eq. (4) into a Fourier series to demonstrate this result analytically, following the previous literature[12], and we express the result in Cartesian coordinates to make the phase structure explicit:

$$\vec{E}(R_0,\theta,\phi,\omega t) = \mathrm{Re}\sum_{l=1}^{\infty}\frac{e}{cR}l\omega\{\varepsilon_+^l(\theta)e^{i(l-1)\phi}\vec{e}_+ + \varepsilon_-^l(\theta)e^{i(l+1)\phi}\vec{e}_- + i\varepsilon_z^l(\theta)\vec{e}_z e^{il\phi}\}e^{-il(\omega t - \frac{R_0}{c})}$$

$$\varepsilon_\pm^l(\theta) \equiv \frac{\varepsilon_x^l(\theta) \pm \varepsilon_y^l(\theta)}{\sqrt{2}} = \beta J_l'(l\beta\sin\theta) \pm \frac{\cos^2\theta}{\sin\theta}J_l(l\beta\sin\theta) \qquad (5)$$

$$\varepsilon_z^l \equiv \cos\theta J_l(l\beta\sin\theta)$$

where $J_l$ and $J_l'$ are a Bessel function of the first kind and its derivative, respectively. Here, we have introduced rotation vectors related to the unit vectors in the x and y directions as $\vec{e}_\pm = (\vec{e}_x \pm i\vec{e}_y)/\sqrt{2}$. The first term in the parentheses represents circular polarised components with the same helicity as that of the electron motion, and the second term represents components with the reverse helicity. The third term arises from the spherical nature of the field. The summation of these three terms represents the elliptical polarisation. The electric fields of the fundamental, second and third harmonics calculated from Eq. (5) are shown in Figure 3. The vortex nature is clearly observed in the electric field distribution of the second and third harmonics but not in the electric field distribution of the fundamental component.

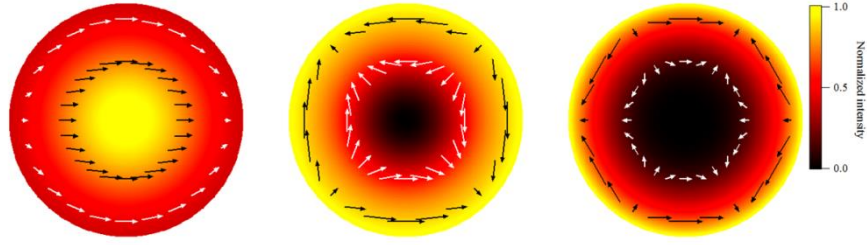

Figure 3. Electric field distribution in the upper hemisphere viewed from the z-direction (see Figure 1), from left to right, for the fundamental ($l=1$), second ($l=2$) and third ($l=3$) harmonics calculated from Eq. (5). The colour represents the field intensity. The fundamental frequency has an intensity maximum in the centre, whereas the harmonics show zero intensity at the centre. Arrows represent the direction of the electric field at a specific time.

In Eq. (5), when $\theta$ is small, the first term becomes dominant and the field is accurately represented by paraxial approximation. Its polarisation is circular and has a phase term $\exp\{i(l-1)\varphi\}$ that is a common feature of a vortex beam. According to Allen et al.[1], such a field possesses an orbital angular momentum of $l-1$. In the general case of $\theta$, the field represented by Eq. (5) is non-paraxial. It has been argued that the radiation field of such a case possesses angular momentum, even though the spin and orbital angular momentum are difficult to separate[14]. We have successfully derived an expression for the ratio of the angular momentum density to the energy density for the radiation emitted by an electron in circular motion, which will be presented in a separate paper. The expression shows that a photon of the $l$-th harmonic carries the total angular momentum whose z-component is equal to $l$. We speculate that the first component in Eq. (5) has spin angular momentum of $+1$ and orbital angular momentum of $l-1$, whereas the spin and orbital angular momenta of the second component are $-1$ and $l+1$, respectively. The total angular momentum in the z direction is always $l$.

When an electron in circular motion drifts along the *z*-axis with high relativistic velocity, the radiation field perpendicular to the *z*-axis is strengthened by the Lorentz factor $\gamma$, which is given by the electron energy divided by the electron rest mass energy, and is collimated into a narrow cone around the *z*-axis[12]. Consequently, the field is well represented by the paraxial approximation. Such radiation fields can be produced in the laboratory using a helical undulator. These devices are widely used as synchrotron light sources[15], wherein a high-energy electron beam executes spiral motion in a specially designed magnetic field, radiating circularly polarised light. This has been identified as an OAM photon source both mathematically[16] and experimentally[17], although the origin of the twisting has not been addressed. The helical undulator radiation corresponds to the case where an electron in circular motion travels towards the $z$-axis at relativistic velocity. Because the phase is Lorentz invariant[18], the harmonic components of the helical undulator radiation should preserve the helical phase structure, which is consistent with the conservation of angular momentum along the direction of motion of the frame in the Lorentz transformation[12].

We have experimentally investigated helical undulator radiation using various methods. Here, we present selected results that provide clear evidence of the twisted nature of the radiation. First, we discuss the interference between the fundamental radiation and harmonics. This result is demonstrated for the fundamental and second harmonics[17]. We present cases including the higher harmonics for the first time. Fig. 4 shows the interference patterns between the fundamental radiation from one undulator and the second or third harmonics from another undulator. As expected from the analytic calculation[17], single- and double-spiral structures are clearly observed. Moreover, the directions of the spiral structures are reversed with the reversal of the electron circulation direction. These results indicate that the relative phase difference between the light beams is in agreement with the theoretical prediction of Eq. (5). However, at this stage, the absolute phase structure of radiation generated by a single undulator is unclear.

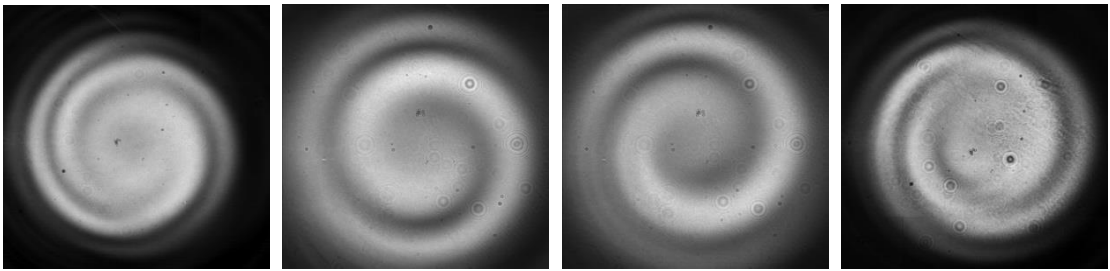

Figure 4. Interference patterns between two undulator radiations. From left to right: patterns between the fundamental and third harmonics and between the fundamental and second harmonics for left-handed polarisation and between the fundamental and second harmonics and between the fundamental and third harmonics for right-handed polarisation.

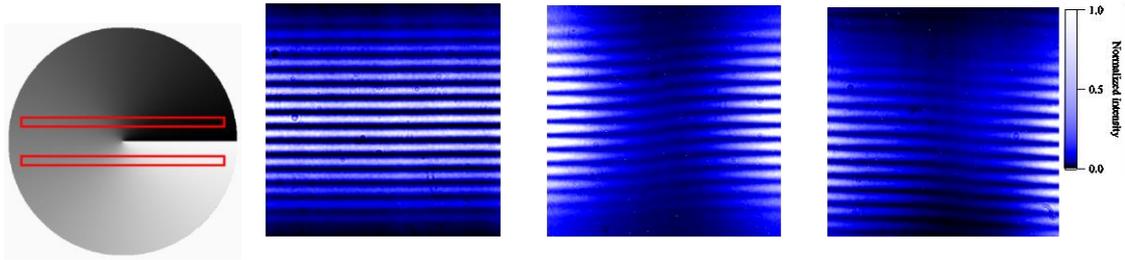

Figure 5. From left to right: Principle of the double-slit diffraction of twisted radiation and the double-slit diffraction pattern of the fundamental radiation, second harmonic radiation (right-handed circular polarisation) and second harmonic (left-handed circular polarisation). In the leftmost illustration, the darkness/brightness indicates the phase, and the red rectangles represent the slits. In the three images showing the measured data, the darkness/brightness indicates the intensity.

Next, we present the results from a double-slit diffraction experiment. This is the first experimental result to demonstrate the vortex nature of radiation from a single undulator. The fundamental and second harmonic components in the ultraviolet range were extracted and irradiated onto a double-slit apparatus. The diffraction patterns were observed using a charge-coupled device (CCD) camera, as shown in Fig. 5. An ordinary stripe pattern was observed for the fundamental radiation, as expected for plane waves. However, for the second harmonic, we observed a singularity in the middle of the pattern, as has been demonstrated for a Laguerre–Gaussian beam created by a laser[19]. This singularity arises from the change in the phase difference between the two slits, as illustrated in Fig. 5. The singularity appeared only when the centre of the beam was located between the slits. The observed patterns were accurately reproduced by SRW simulation code [20]. These results clearly show that the harmonics exhibit a twisted phase structure, whereas the fundamental component does not.

We have shown that the harmonic components of an electromagnetic field radiated by electrons in circular motion naturally have a helical phase structure, which suggests the presence of orbital angular momentum. We demonstrated this experimentally by observing helical undulator radiation. The generation of a twisted photon beam from a helically micro-bunched electron beam was reported[21]. However, a vortex radiation field was formed by the constructive interference of the non-twisted radiation from the electrons that were helically aligned in space. This is analogous to the generation of monochromatic radiation from a micro-bunched beam travelling in a uniform magnetic field, even though each electron produces broadband synchrotron radiation[22].

This work allows us to predict the conditions under which twisted radiation will be produced. We propose that cyclotron/synchrotron radiation, particularly from an electron cyclotron maser[23], should be re-examined as twisted radiation. Another candidate is the non-linear inverse Compton scattering[24] of circular-polarised light. In this case, the intense incoming light field causes relativistic circular motion of electrons, thereby producing twisted harmonic radiation. These radiation processes may play important, unexplored roles in the solar magnetosphere[10] or planetary magnetosphere[25], around magnetised neutron stars[26], around active galactic nuclei[27], in tokamaks used for nuclear fusion[28], or in particle accelerators[15]. Observations using phase information may provide new approaches to the analysis of such systems. Several methods have been proposed for detecting OAM photons in nature, but there has been no explicit discussion of the sources of such radiation[5,6,29].

This work also suggests possible technologies for producing twisted photon beams in laboratories at wavelengths ranging from radio waves to gamma rays. In the ultraviolet and X-ray ranges, helical undulators can provide high-brightness twisted radiation. In the microwave and terahertz ranges, gyrotrons, where high-energy electrons execute circular motion and produce cyclotron radiation with harmonics, are potentially powerful sources of twisted radiation[30]. Another candidate is the non-linear inverse Compton scattering[24] of intense circular-polarised laser radiation by a relativistic electron beam provided by an accelerator, which could potentially provide a twisted X-ray or gamma-ray source. The development of these light sources will expand the application of twisted radiation to the entire electromagnetic wavelength range.

ACKNOWLEDGEMENT

Part of this work was supported by JSPS KAKENHI Grant Numbers 26286081 and 26390112, and by the Quantum Beam Technology Program of MEXT/JST. We appreciate the IMS Equipment Development Center for providing the double slits. We thank to Profs. S. Kubo, Y. Yoshimura, K. Ohmi, and T. Saito and Drs. K. Tsuchiya and T. Hayakawa for their helpful discussions. MK contributed to all the theoretical work. TK, MH, YTai, YTak, AMo, and MF verified the analytic calculations and contributed to the execution of the numerical calculations. MK and MH contributed to all experiments. SS, TK, NY, Ami, KM and KK contributed to the interference experiment. MF and NSM contributed to the diffraction experiments. The manuscript was prepared by MK and was checked by all other authors. The current affiliation of NY and TK is the High Energy Accelerator Research Organization (KEK).

APPENDIX

This experiment was conducted at the BL1U beamline of the UVSOR-III electron storage ring, which is equipped with two polarisation-variable undulators in tandem. The undulator had 10 magnetic periods with an 88-mm period length. The variable-polarisation undulator was operated in circular-polarised mode during the experiments. In the undulators, the electron beam executes spiral motion, thereby producing quasi-monochromatic synchrotron radiation and harmonics in the ultraviolet wavelength range. This major advantage of our experiment allowed us to perform all of the experiments in the air using ordinary optical components and devices. The UVSOR-III electron storage ring was operated at 500 MeV, lower than the nominal electron energy of 750 MeV. The electron beam emittance at this energy was estimated to be 8 nm-rad. The electron beam was diffraction-limited in the UV wavelength range; therefore, the undulator radiation was spatially coherent, which was essential for the experiments described below. The typical electron beam current used in the experiment was 1 mA, much smaller than the normal operating current but sufficient for these measurements. The undulator radiation was extracted from the accelerator through the $SiO_2$ window, without the use of mirrors or monochromators.

In the interference experiment, two undulators in tandem were tuned to produce either the fundamental radiation or harmonics at the same wavelength, i.e., 355 nm. The two interfered light beams were observed via a CCD camera with a bandpass filter at 355 nm and a width of 1.3 nm. In the diffraction experiments, a double slit with a width of 200 microns and separation of 2 mm was positioned approximately 7.5 m from the centre of the helical undulator and was aligned so that the optical beam centre was in the middle of the slits. The diffraction image was observed via a CCD camera with a 355-nm bandpass filter placed 3 m downstream from the slit. A diffraction pattern with a singularity, as shown in Fig. 6, appeared when the beam centre was located between the slits but not when the slit position was shifted vertically by 3 mm in either direction. To obtain a clear diffraction pattern for the images presented in this paper, a linear polarised filter was placed in the vertical direction to eliminate contamination from the horizontally polarised radiation caused by the bending magnets. Neutral density (ND) filters were added as necessary to prevent saturation of the CCD camera.

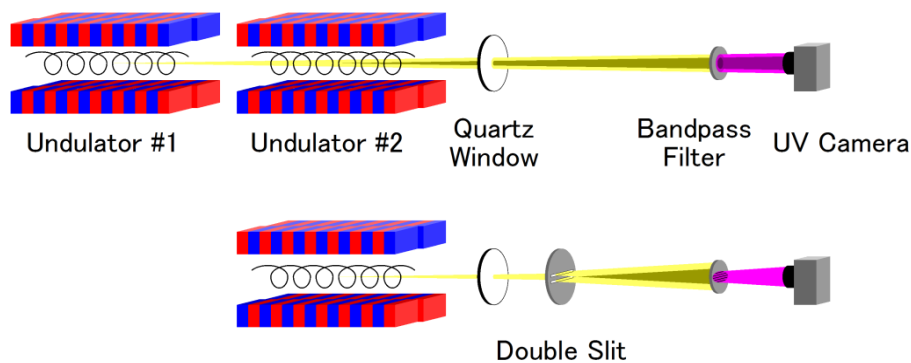

Figure 6. Experimental setup of the interference experiment (upper) and the double-slit diffraction experiment (lower). Electrons travel from left to right while executing spiral motion.

**References;**

1. L. Allen, M. W. Beijersbergen, R. J. C. Spreeuw, J. P. Woerdman, *Phys. Rev. A*. **45**, 8185–8189 (1992)

2. J. P. Torres, L. Torner (eds.) *Twisted Photons; Application of Light with Orbital Angular Momentum*, WILEY-VCH Verlag GmbH & Co. KGaA (2011)

3. G. Molina-Terriza, J. P. Torres, L. Torner, *Nat. Phys*. **3**, 305–310 (2007)

4. M. van Veenendaal, I. McNulty, *Phys. Rev. Lett*. **98**, 157401 (2007)

5. M. Harwit, M. *Astrophys. J*. **597**, 1266–1270 (2003)

6. N. M. Elias II, *Astron. & Astrophys*. **492**, 883–922 (2008)

7. M. D. Gray, , G. Pisano, S. Maccalli, P. Schemmel, *Mon. Not. R. Astron. Soc*. **445**, 4477–4503 (2014)

8. F. Tamburini, B. Thidé, G. Molina-Terriza, G. Anzolin, *Nat. Phys*. **7**, 195–197 (2011)

9. O. Heaviside, O. *Nature*. **69**, 293 (1904)

10. T. Takakura, *Solar Phys*. **1**, 304–353 (1967)

11. J. Schwinger, *Phys. Rev*. **75**, 1912–1925 (1949)

12. L. D. Landau, E. M. Lifshitz, *The classical theory of fields*. (4th Rev. English Ed.), Elsevier Ltd. (1975)

13. J. D. Jackson, *Classical Electrodynamics*. (3rd ed.) John Wiley & Sons, Inc. (1999)

14. S. M. Barnett, L. Allen, *Opt. Comm.* **110**, 670–678 (1994)

15. H. Winick, H. (ed.) *Synchrotron Radiation Sources; A Primer*. World Scientific Publishing Co. Pte Ltd (2014)

16. S. Sasaki, I. McNulty, *Phys. Rev. Lett*. **100**, 124801 (2008)

17. J. Bahrdt, *Phys. Rev. Lett*. **111**, 034801 (2013)

18. A. Einstein, *Annalen der Physik*. **17**, 891–921 (1905)

19. H. L. Sztul, R. R. Alfano, *Opt. Lett*. **31**, 999–1001 (2006)
20. O. Chubar, *Nucl. Instr. Meth. Phys. Res.* A 435, 495-508 （1999）
21. E. Hemsing, *et al*. *Nat. Phys*. **9**, 549–553 (2013).
22. S. Bielawski, S. *et al*. *Nat. Phys.* **4**, 390–393 (2008)
23. R. A. Treumann, *Astron. Astrophys. Rev*. **13**, 229–315 (2006)
24. D. D. Meyerhofer, *IEEE J. Quantum Electron*., 33(11), 1935 (1997)
25. I. de Pater, I. *Adv. Space Res*. **3**, 3l–37 (1983)
26. V. L. Ginzburg, S. I. Syrovatskii, S. I., *Annu. Rev. Astron. Astrophys*. **3**, 297–350 (1965)
27. H. Krawczynski, E. Treister, *Front. Phys.* **8**, 609–629 (2013)
28. M. Bornatici, M. *et al.*, *Nucl. Fusion*, 23(9), 1153 (1983)
29. G. C. G. Berkhout, M. W. Beijersbergen, *Phys. Rev. Lett.* 101, 100801 (2008)
30. Kartikeyan, M. V. *et al.*, "Gyrotrons: High-Power Microwave and Millimeter Wave Technology" (Advanced Texts in Physics) Springer (2004)